\documentclass[12pt,english]{article}
\usepackage[T1]{fontenc}
\usepackage[latin9]{inputenc}
\usepackage{geometry}
\usepackage{float}
\usepackage{textcomp}
\usepackage{amstext}
\usepackage{graphicx}
\usepackage{setspace}
\usepackage[authoryear]{natbib}
\makeatletter

\providecommand{\tabularnewline}{\\}

\@ifundefined{showcaptionsetup}{}{%
 \PassOptionsToPackage{caption=false}{subfig}}
\usepackage{subfig}
\makeatother

\usepackage{babel}
\begin{document}

\title{Topography of (exo)planets}

\author{F. Landais (1{*}), F. Schmidt (1), and S. Lovejoy (2)}
\maketitle

\paragraph*{Affiliations:}

(1) GEOPS, Univ. Paris-Sud, CNRS, Universit\'{e} Paris Saclay, Rue du
Belvedere, Bat. 504\textendash 509, 91405 Orsay, France (2) Physics
department, McGill University, 3600 University st., Montreal, Que.
H3A 2T8, Canada 

\paragraph*{{*}Correspondence to: }

F. Landais (francois.landais@u-psud.fr)
\begin{abstract}
Current technology is not able to map the topography of rocky exoplanets,
simply because the objects are too faint and far away to resolve them.
Nevertheless, indirect effect of topography should be soon observable
thanks to photometry techniques, and the possibility of detecting
specular reflections. In addition, topography may have a strong effect
on Earth-like exoplanet climates because oceans and mountains affect
the distribution of clouds \citep{Houze2012}. Also topography is
critical for evaluating surface habitability \citep{Dohm2015}.

We propose here a general statistical theory to describe and generate
realistic synthetic topographies of rocky exoplanetary bodies. In
the solar system, we have examined the best-known bodies: the Earth,
Moon, Mars and Mercury. It turns out that despite their differences,
they all can be described by multifractral statistics, although with
different parameters. Assuming that this property is universal, we
propose here a model to simulate 2D spherical random field that mimics
a rocky planetary body in a stellar system. We also propose to apply
this model to estimate the statistics of oceans and continents to
help to better assess the habitability of distant worlds.
\end{abstract}
Keywords: planetary systems, planets and satellites: surfaces, planets
and satellites: terrestrial planets, methods: numerical

\section{Introduction}

Efforts to detect and study exoplanets in other solar systems were
initially restricted to gas giants \citep{Mayor1995} but multiple
rocky exoplanets have now been discovered \citep{Wordsworth2011}.
Their climates depend mainly on their atmospheric composition, stellar
flux and orbital parameters \citep{Wang2014,Forget2014}. But topography
also plays a role in atmospheric circulation \citep{Blumsack1971}
and is an important trigger for cloud formation \citep{Houze2012}.
Furthermore, the presence of an ocean filled with volatile compounds
at low albedo is of a prime importance to the climate \citep{Charnay_Exploringfaintyoung_JGRA2013}.
Last but not least, surface habitability relies on the presence of
the three elements: the atmosphere, ocean and land \citep{Dohm2015}.
Topography is also the determinant of ocean and land cover. 

Thanks to different observations techniques, measurements of the atmospheres
of hot Jupiter planets have been achieved \citep{Seager2010}. Significantly,
the detection of clouds has been reported \citep{Demory2013} indicating
strong heterogeneity in their spatial distribution. The detection
of the first atmospheric transmission spectra of a super-Earth \citep{Bean2010}
and the discovery of a rocky exoplanet in the habitable zone around
a dwarf star opens a new area in exoplanet science \citep{Wit2016}.
Such observations are expected to be increasingly frequent \citep{Ti2015}.
Nevertheless, with current technology, direct imaging of exoplanets
is very difficult because the objects are too faint and too far away.
For the moment, the only way to determine the topography is by statistical
models. 

In the near future, photometry techniques should improve our knowledge
of exoplanet topography , even if the bodies are not resolved in ways
similar to the small bodies in our Solar System (see for instance
\citet{lowry2012nucleus}  for estimates of the shape of comet 67P
before the Rosetta landing). In addition, if oceans or lakes are present,
their specular reflection should be detectable , for example, as also
observed through the haze of Titan \citep{stephan2010specular} .
Even if exoplanets are too far to be resolved, their topographies
should be studied now. We offer here a framework to prepare and interpret
future observations.

Recently, we reported the first unifying statistical similarity between
the topographic fields of the best known bodies in the Solar System:
Earth, Moon, Mars and Mercury \citep{Landais2018}. All these topographies
seems to be well described by a mathematical scaling framework called
«multifractals». The multifractal model, initially proposed for topography
by \citet{Lavallee1993} describes the distribution and correlation
of slopes at different scales. More precisely, we consider here the
``universal multifractal'' model developed by \citet{Schertzer1987}.
The accuracy of such a model has been tested in the case of different
available topographic fields on Earth \citep{Gagnon2006}, Mars \citep{Landais2015},
Mercury and the Moon \citep{Landais2018}. This model has the advantage
to reproduce closely the statistical properties of natural topography:
the scaling properties, but also the intermittency (both rough and
smooth regions can be found on the planets). Universal multifractals
depend on only 3 parameters: $H$ controls how the roughness changes
from one scale to another and $C_{1}$ controls the spatial heterogeneity
of the roughness near the mean and $\alpha$ quantifies how rapidly
the properties change as we move away from the mean topographic level.
The bodies studied show transitions at \textasciitilde{}10 km and
are characterized by specific multifractal parameters \citep{Landais2018}.
The scaling law at large scales (> 10 km) is characterized for the
Moon by $H=0.2$, Mercury by $H=0.3$, Mars and Earth by $H=0.5$.
The $\alpha\sim1.9$ for the Earth, Mars and Mercury but $\alpha\sim1.4$
for the Moon. The $C_{1}\sim0.1$ for Earth and Mars, with lower values
$C_{1}\sim0.06$ for Mercury and $C_{1}\sim0.03$ for the Moon. These
differences are interpreted to be linked to dynamical topography and
variation of elastic thickness of the crust (see table \ref{tab:Estimates-of-the}).

Assuming that exoplanets are statistically similar to those observed
in our own Solar System, we propose here a stochastic topographic
model. Such models will be very useful for investigating the distribution
of exoplanet oceans, for studying the effect of topography on exoplanet
climates, and for studying the effect of topography on their orbital
motions or for determining the effect of topography and roughness
on photometry. It can also be used to study the early climate on Earth.

The purpose of this article to first present our statistical model
and its implementation on the sphere we then discuss the distribution
of oceans and land cover. An introduction to the multifractal formalism
can be found in the next section.

\section{Method}

\subsection{Universal multifractals\label{subsec:Universal-multifractals}}

The first application of fractional dimensions on topography was
by B. Mandelbrot in his article ``how long is the coast of Britain''
\citep{Mandelbrot1967}. Fractals are geometrical sets of points that
have scaling, power law, deterministic or statistical relations from
one scale to another. This type of behavior has been observed in geophysical
phenomenon including turbulence - clouds, wind, ocean gyres - but
also faults in rock, geogravity, geomagnetism and topography \citep{lovejoy2007scaling}

. The most common way to test scaling is to study the dependence of
various statistics as functions of scale. Topographic level contours
(isoheights) are \emph{fractals} if for example the length of the
contour is a power law function of the resolution at which it is measured.
In this case, the level set is ``scaling'' and the exponent is its
fractal dimension. In real topography, each level set has its own
different fractal dimension so that the topography itself is a multifractal
\citep{Lavallee1993}. Numerous studies haves shown that in several
contexts, topography is scaling over a significant range of scales
(see the review in \citealp{Lovejoy2013a}). If the topography is
multifractal, fractal dimensions measured locally appear to vary from
one location to another. Indeed multifractal fields can be thought
as a hierarchy of singularities whose exponents are random variables.
Modern developments have introduced the notion of \emph{multifractal}
processes for such fields. For such processes, a local estimate of
a fractal exponent is expected be different from a location to another
without requiring different processes to generate it. With multifractals,
it is possible to interpret the topography of regions that exhibit
completely different slope distributions in a unified statistical
framework. These models suggest global topography analyses are relevant
despite of their diversity and complexity. Previous studies \citep{Gagnon2006,Lavallee1993}
have established the accuracy of multifractal global statistical approach
in the case of Earth's topography. More precisely, a particular class
of multifractal has been considered: the universal multifractal, a
stable and attractive class \citep{Schertzer1987}. In our previous
analysis \citep{Landais2015}, we performed the same kind of global
analysis on the topographic data from Mars, from MOLA laser altimeter
measurement \citep{Smith2001}. This analysis also find a good agreement
with universal multifractal but on a restricted range of scale \citep{Landais2015}.
Indeed the statistical structure has been found to be different at
small scale (monofractal) and large scale (multifractal) with a transition
occurring around 10 km. 

\paragraph{Fluctuations}

In order to interpret topography as a multifractal, we must quantify
its fluctuations. The simplest fluctuation that can be used to describe
topography is the distribution of changes in altitude $\Delta h$
over horizontal distances $\Delta x$. There are many other ways to
define fluctuations, the general framework being wavelets. The simple
altitude difference corresponds to the so called ``poor man's''
wavelet and can be efficiently replaced by the Haar wavelet that tends
to converge faster and is useful over a wider range of geophysical
process. Over an interval $\Delta x$, the Haar fluctuation is the
average elevation over the first half of the interval minus the average
elevation over the second half (see \citealp{Lovejoy2014,Lovejoy2012})
and paragraph below for a precise definition of Haar fluctuations).
The computation of fluctuations can be performed for each pair of
elevation data in order to accumulate a huge amount of slope fluctuations.
From this, a global planetary average $M(\Delta x)$ can be performed
and will reflect the mean fluctuation of slopes at the scale $\Delta x$. 

\paragraph{Scaling}

By estimating fluctuations at different scales, we can observe the
structure of the statistical dependance of the ensemble mean fluctuation
at scale $\Delta x$: $M(\Delta x)$ . If the topographic field is
fractal, this dependance is a power-law corresponding to equation
\ref{eq:Moment1} where H is a power law exponent (named in Honor
of Ewin Hurst and equal to the Hurst exponent in the monofractal,
Gaussian case):

\begin{equation}
M(\Delta x)\sim\Delta x^{H}\label{eq:Moment1}
\end{equation}

\paragraph{Statistical moments}

Additionally, instead of simply considering the average (i. e.the
first statistical moment of the fluctuations), we can compute any
statistical moment $M_{q}$ of order $q$ defined by $M_{q}=<\Delta h^{q}>$
; $M_{q}$ is called the $q^{\mathrm{th}}$ order structure function.
If $q=2,$ it simply corresponds to the usual (variance based) structure
function. In principle, all orders (including non-integer orders)
must be computed to fully characterize the full variability of the
data. 

\paragraph{Multifractality}

$M_{q}$ allows us to introduce two distinct statistical structures
of interest: monofractal and multifractal. For a detailed description
of the formalism we apply in this study, the readers can refer to
\citet{Lovejoy2013a} briefly summed in \citet{Landais2015}. We quickly
recall the main notions here :
\begin{itemize}
\item In the monofractal case the parameters $H$ is sufficient to describe
the statistics of all the moments of order $q$ (equation \ref{eq:monfractal}).
In this case, no intermittency is expected, meaning that the roughness
of the field is spatially homogenous despite of its fractal variability
regarding to scales. Typically, the value $H=0.5$ corresponds to
the classic Brownian motion. This kind of model has been used in many
local and regional analysis of natural surfaces \citep{Orosei2003,Rosenburg2011},
but it fails to account for the intermittency (and strongly non-Gaussian
statistics) commonly observed on large topographic datasets. 
\begin{equation}
M_{q}\sim\Delta x^{qH}\label{eq:monfractal}
\end{equation}
\item In the multifractal case, $H$ is no longer sufficient to fully describe
the statistics of the moments of order $q$. An additional convex
function $K(q)$ depending on $q$ is required (\ref{eq:multifractal}).
\begin{equation}
M_{q}\sim\Delta x^{qH-K(q)};\label{eq:multifractal}
\end{equation}
\item The moment scaling function $K$ modifies the scaling law of each
moment. The consequence on the corresponding field appears clearly
on simulations: the field exhibit a juxtaposition of rough and small
places that are clearly more realistic in the case of natural surfaces
\citep{Gagnon2006}. Moreover, it is possible to restrain the generality
of the function $K(q)$ by considering universal multifractals, a
stable and attractive class proposed by \citet{Schertzer1987} for
which the multifractality is completely determined by the mean intermittency
$C_{1}=\left(\frac{dK(q)}{dq}\right)_{q=1}$ (codimension of the mean)
and the curvature $\alpha$ of the function $K$, $\alpha=\frac{1}{C_{1}}\frac{d^{2}K(q)}{dq^{2}}$
evaluated at $q=1$ (the degree of multifractality). In this case
the expression of $K$ is simply given by equation \ref{eq:universal}
\end{itemize}
\begin{equation}
K(q)=\frac{C_{1}}{\alpha-1}(q^{\alpha}-q)\label{eq:universal}
\end{equation}

\subsection{Spherical multifractal simulation\label{subsec:Spherical-multifractal-simulatio}}

Simulations in 1D or 2D with multifractal properties and specific
values for $\alpha$, $H$ and $C_{1}$ can be obtained by the procedure
defined by \citet{Schertzer1987,Wilson1991}. The necessary steps
are briefly reminded here after :
\begin{itemize}
\item Step 1 : Generation of a un-correlated Levy noise $\gamma_{\alpha}(r)$.
When $\alpha=2,$ it simplifies to a gaussian white noise whereas
$\alpha<2$ corresponds to an extremal levy variable with negative
extreme values. 
\item Step 2 : Convolution of $\gamma_{\alpha}(r)$ with a singularity $g_{\alpha}(r)$
defined by equation \ref{eq:singularity} to obtain a Levy-generator
$\Gamma_{\alpha}(r)$, by using a convolution denoted by ``$\star"$
\begin{equation}
g_{\alpha}(r)=|r|^{-2/\alpha}\label{eq:singularity}
\end{equation}
\begin{equation}
\Gamma(r)=C_{1}^{1/\alpha}g(r)\star\gamma_{\alpha}(r)\label{eq:singularity2}
\end{equation}
\item Step 3 : Exponentiation of the generator to obtain the multifractal
noise $\varepsilon$
\begin{equation}
\varepsilon=e^{\Gamma}\label{eq:MultifractalNoise}
\end{equation}
\item Step 4 : The final field is then obtained by fractional integration
of order H (another convolution similar to step 2)
\end{itemize}
Whereas the convolutions required for step 2 and 4 can easily be performed
in Fourier space for the cartesian case, the generalization to spherical
case is not straightforward, but as shown in appendix 5D of \citet{Lovejoy2013a},
it can be done using spherical harmonics. Let $\theta$ and $\varphi$
being respectively the colatitude and longitude angle, the singularity
can be expressed by equation \ref{eq:singularitySPHERIC}. As it is
symmetric by rotation along $\varphi$, $g_{\alpha}(\theta,\varphi)$
only depend on $\theta$. 

\begin{equation}
g_{\alpha}(\theta,\varphi)=\theta^{-2/\alpha}\label{eq:singularitySPHERIC}
\end{equation}

Let the spherical harmonic expansion of $g_{\alpha}(\theta,\varphi)$
be given by equation \ref{eq:decomp}, where $Y_{lm}$ is the spherical
harmonic of order $m$ and $l$. As $g_{\alpha}(\theta,\varphi)$
does not depend on $\varphi$, all the $Y_{lm}$ for $m\neq0$ are
equal to zero.

\begin{equation}
g_{\alpha}(\theta,\varphi)=\sum\sigma_{l}Y_{l,0}\label{eq:decomp}
\end{equation}

Let the spherical harmonic expansion of $\gamma_{\alpha}(\theta,\varphi)$
be given by :

\begin{equation}
\gamma_{\alpha}(\theta,\varphi)=\sum u_{lm}Y_{l,m}(\theta,\varphi)
\end{equation}

Then the convolution $C$ of $g_{\alpha}(\theta,\varphi)$ and $\gamma_{\alpha}(\theta,\varphi)$
is given by :

\begin{equation}
C=\sum_{l,m}\sigma_{l}\sqrt{\frac{4\pi}{2l+1}}u_{lm}Y_{l,m}(\theta,\varphi)
\end{equation}

\section{Results}

\subsection{Solar System}

\begin{figure}
\begin{centering}
\includegraphics[width=1\textwidth]{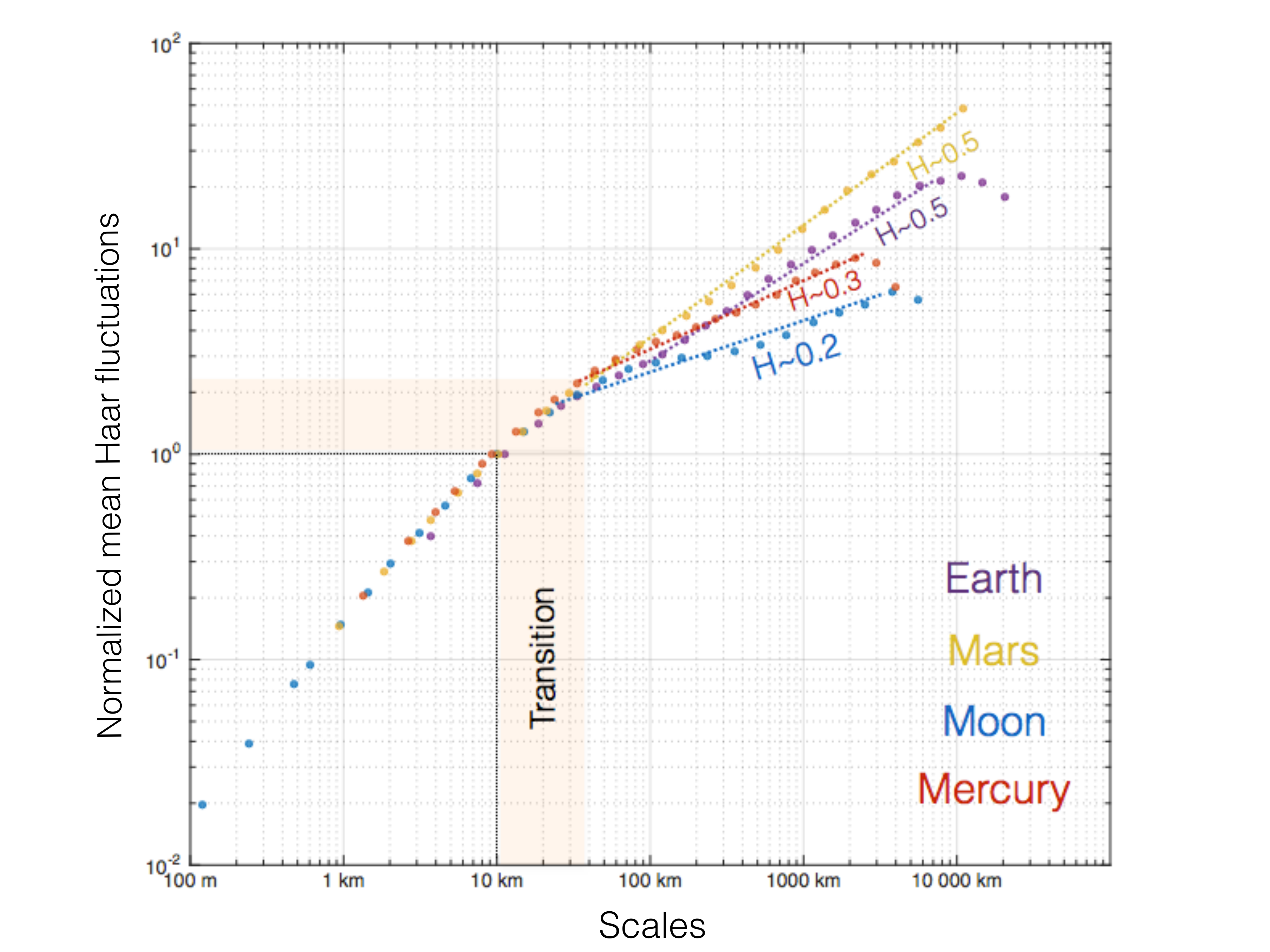}
\par\end{centering}
\caption{Mean fluctuations of topography of Earth, Mars, Moon and Mercury,
as a function of scale. All dataset are normalized in order to be
equal to 1 at the scale 10km. The normalization does not modify the
scaling behavior but emphasize the transition occurring at around
10km. The errors bars are smaller than the size of the points. \label{fig:Mean-fluctuations-SolarSytem}}
\end{figure}

In this section, we recall the main results of the planetary bodies
of the Solar System. On Figure \ref{fig:Mean-fluctuations-SolarSytem},
we have plotted the mean normalized fluctuations of altitude as a
function of scale on a log-log plot. The easiest way to define fluctuations
at a given scale $\Delta x$ is to take the simple difference of altitude
between two points separated by the distance $\Delta x$. We average
all of these fluctuations over the whole planetary body. As we are
focusing on statistical properties, the results on figure \ref{fig:Mean-fluctuations-SolarSytem}
have been normalized in order to emphasize the transition between
2 distinct range of scales. The global average have been normalized
in order to be similar around 10 km. As a consequence of this normalization,
it is not possible to compare the absolute altitude and roughness
values on this plot, only the scaling laws. One can see the similarity
between curves at lower scales (<10 km) and distinct scaling behaviors
at higher scales (>10km). Still in each case, the dependance towards
scales remains roughly linear on a log-log plot revealing a simple
power-law behavior. The parameters $H$ is taken as a function of
the linear coefficient of the fit and thus control how the mean fluctuations
of elevations behave towards scales. This kind of linear behavior
is called fractal or monofractal. 

Moreover the multifractal model includes two other parameters ($C_{1}$
and $\alpha$) that control the spatial distribution of roughness.
Thanks to $C_{1}$ and $\alpha$, it is possible to have a global
description, in a common statistical framework, including regions
with heterogeneous roughness at a given scales. More details about
the two non-trivial parameters may be found in the appendices. Global
measures of $H$, $C_{1}$ and $\alpha$ in the case of Earth, Mars,
Moon and Mercury have produced satisfying results (see table \ref{tab:Estimates-of-the}and
\citealp{Landais2018}). 

We analyzed the generated random field and show that the estimated
$H$, $C_{1}$ and $\alpha$ are in agreement with the expected values
for a large range of parameter space. 

\subsection{Exoplanets}

\begin{table}[H]
\caption{Estimates of the parameters $H$, $\alpha$ and $C_{1}$\label{tab:Estimates-of-the}}

\centering{}%
\begin{tabular}{|c|c|c|c|c|c|c|c|c|}
\hline 
 & \multicolumn{2}{c|}{Earth} & \multicolumn{2}{c|}{Mars } & \multicolumn{2}{c|}{Moon} & \multicolumn{2}{c|}{Mercury}\tabularnewline
\hline 
\hline 
 & low & high & low & high & low & high & low & high\tabularnewline
\hline 
$H$ & 0.8 & 0.5 & 0.7 & 0.5 & 0.9 & 0.2 & 0.7 & 0.3\tabularnewline
\hline 
$C_{1}$ & 0.001 & 0.1 & 0.004 & 0.11 & 0.04 & 0.03 & 0.004 & 0.06\tabularnewline
\hline 
$\alpha$ & NA & 1.9 & NA & 1.8 & NA & 1.4 & NA & 1.9\tabularnewline
\hline 
\end{tabular}
\end{table}

\begin{figure}
\subfloat[Spherical simulations at 0.1\textdegree{} resolution for different
values of $H$ ($\text{\ensuremath{\alpha}=1.9}$ and $C_{1}=0.1$).
$H$ varies from 0.2 to 0.99. Synthetic bodies with low $H$ values
have little large-scale altitude fluctuations and are rough at small
scales. As a result, their shape is similar to a regular sphere but
with a rough texture. When $H$ increases, this behavior tends to
be reversed : large altitude variations appear at large scales deforming
the body, which has a smoother texture.\label{fig:Effect-of-varying-H}]{\includegraphics[width=1\textwidth]{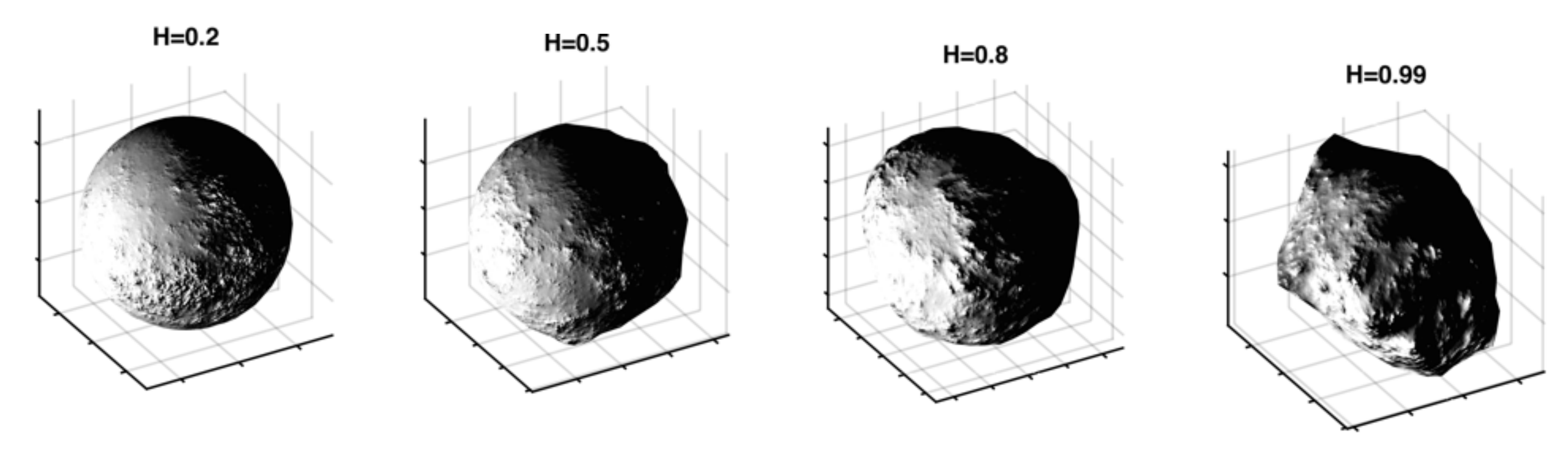}

}

\subfloat[Spherical simulations at 0.1\textdegree{} resolution for two values
of $C_{1}$ ($\text{\ensuremath{\alpha}=1.9}$ and $H=0.5$ constant).
From left to right $C_{1}$ is 0 and 0.1. The left simulation ($C_{1}$
= 0) is characterized by a spatially homogeneous roughness. On the
contrary, themultifractal simulation on the right shows alternating
smooth and rough areas  \label{fig:Effect-of-varying-C1}]{\includegraphics[width=1\textwidth]{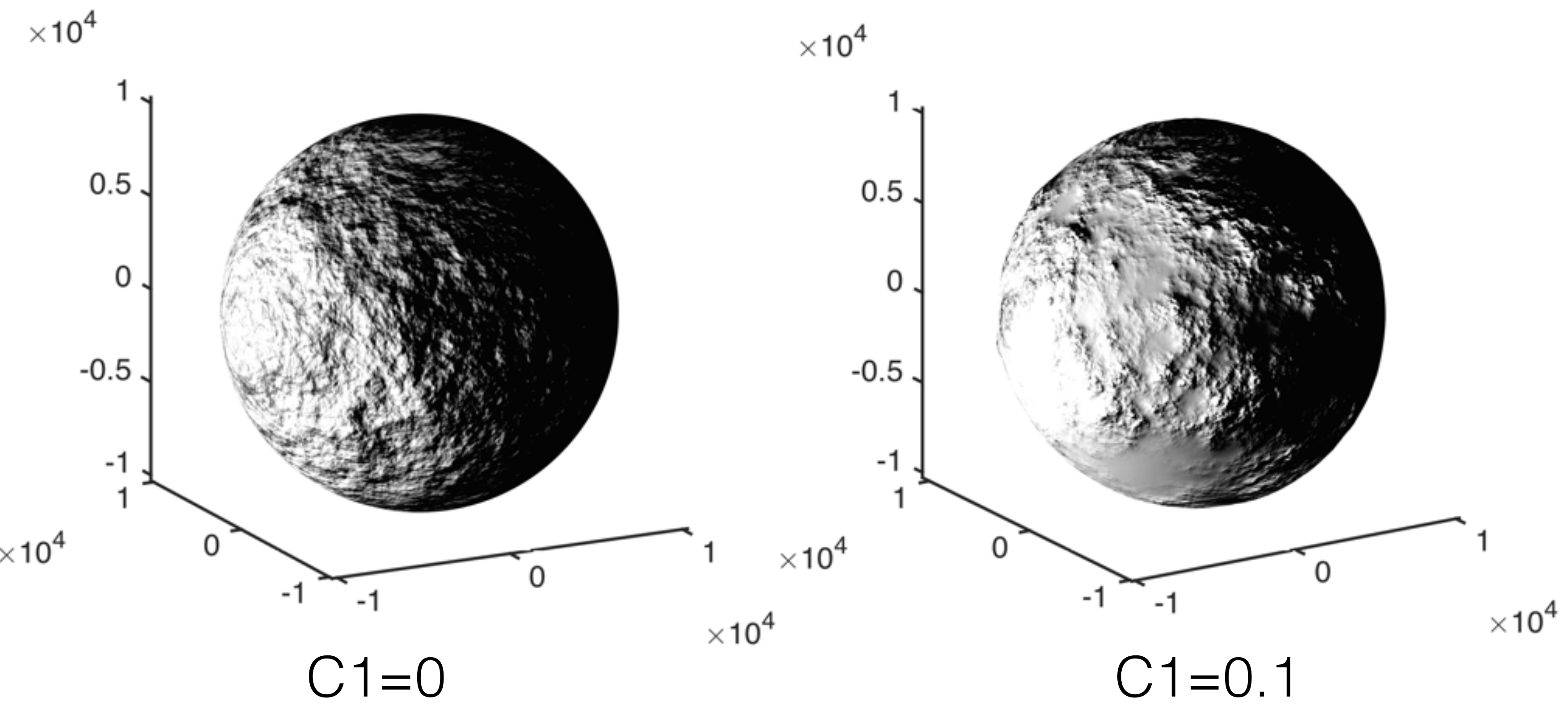}

}

\caption{Several example of synthetic spherical topographic fields by varying
H and $C_{1}$ \label{fig:Several-exemple-of-multifractal-topo}}
\end{figure}

Given its simplicity and its accuracy in the case of several real
topographies, the multifractal model should be a good candidate for
producing artificial topographies of (exo)planets. Figure \ref{fig:Several-exemple-of-multifractal-topo}
provides several examples of spherical topography obtained by our
simulation model for varying values of $C_{1}$ and $H$. One can
see the interesting multifractal features . In the case of non-zero
$C_{1}$, the roughness level is highly heterogeneous with an alternation
of smooth and rough terrains depending on the altitude. This features
makes the multifractal simulations much more realistic by (implicitly)
taking into account the possible occurrence of oceans or large smooth
volcanic plains that are statistically different from deeply cratered
terrains or mountainous areas where the level of roughness is high.
Whereas the value of $H$ controls the rate at which the roughness
changes with scale (see figure \ref{fig:Effect-of-varying-H}), the
value of $C_{1}=0.1$ controls the proportion of rough and smooth
places (see figure \ref{fig:Effect-of-varying-C1}). A high value
increases the roughness discrepancies between locations. One has to
remember that only the scaling laws are simulated here, neither the
absolute height, nor the radius of the planet. Vertical exaggeration
has been set arbitrarily in order to maximize the visual impression.
Nevertheless, the variety of shapes and roughnesses produced are
astonishing and in addition to terrestrial planets, could potentially
even be realistically applied to small bodies including asteroids
and comets.

\begin{figure}
\begin{centering}
\includegraphics[width=0.8\textwidth]{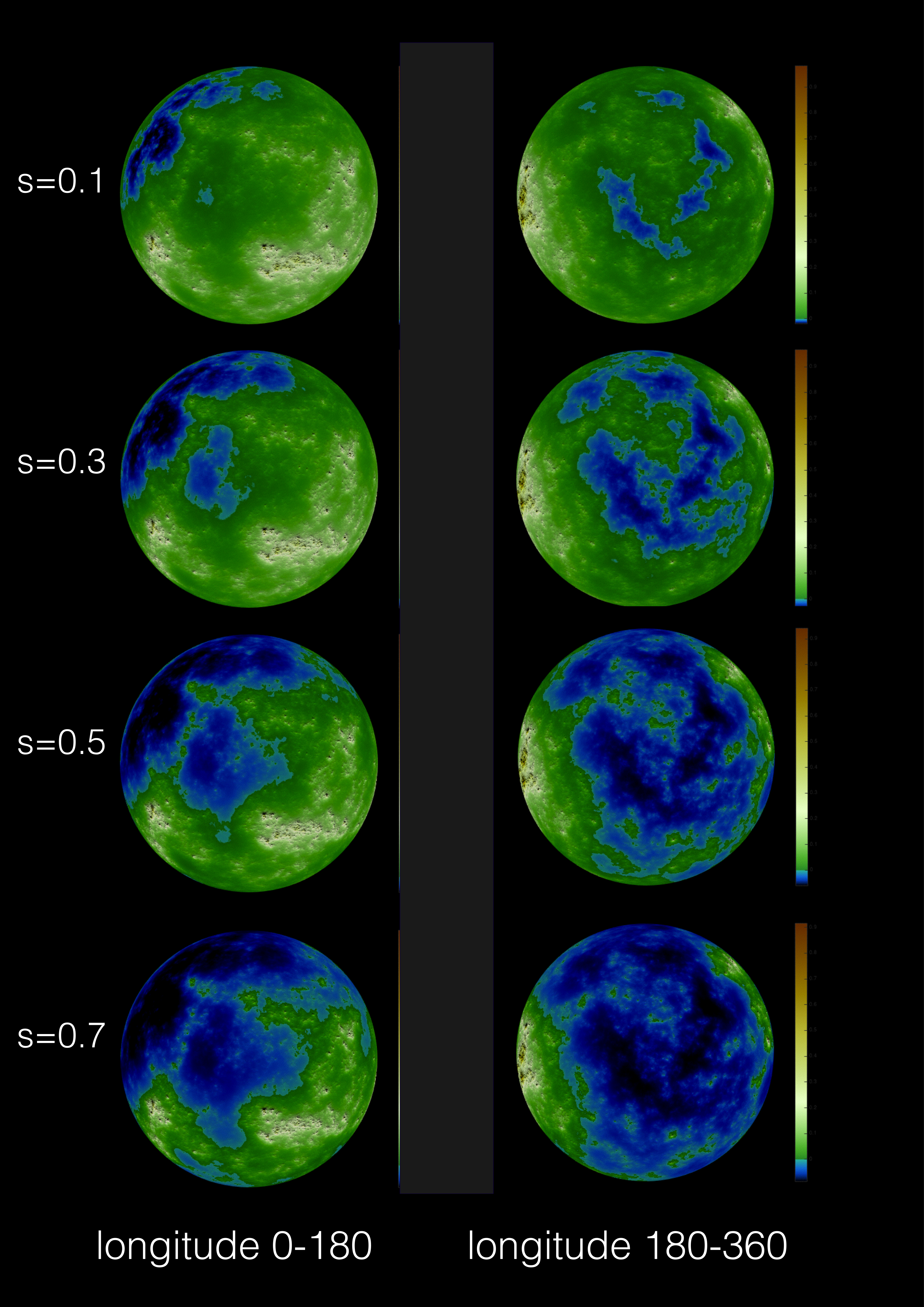}
\par\end{centering}
\caption{Synthetic multifractal topography at 0.1\textdegree{} resolution as
a function of sea level. The fraction of the planet\textquoteright s
surface covered by ocean is noted s. The simulation is set for the
Earth/Mars like planet ($H$ = 0.5, $\alpha$ = 1.9, $C_{1}$ = 0.1).
Low altitude regions are smoother than high altitude ones. See also
video 2 in sup.mat. \label{fig:Example-of-180x360}}
\end{figure}

\begin{figure}
\begin{centering}
\includegraphics[width=1\textwidth]{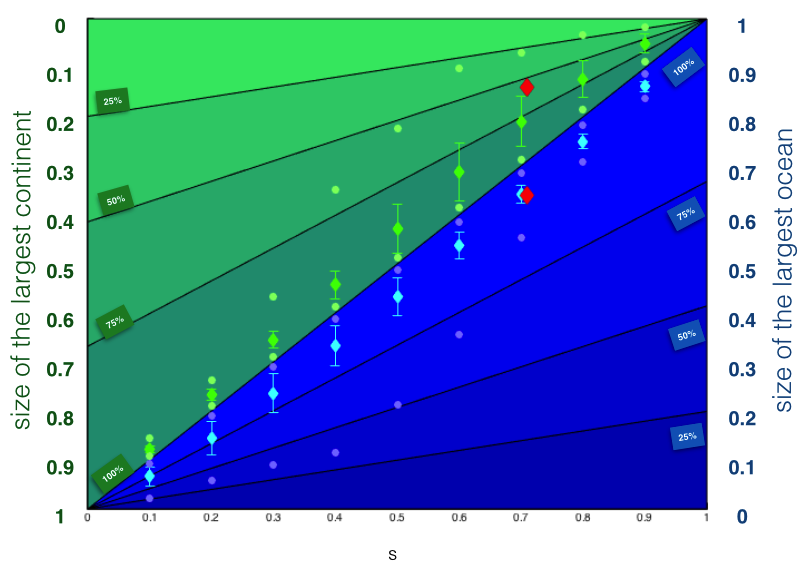}
\par\end{centering}
\caption{The Ocean/continent relationship. The size (as proportion of the total
planet surface) of the largest continent (blue) and ocean (green))
for different values of sea level s. The diamond indicate the mean
size with one standard deviation bars, whereas the circles indicate
the minimum and maximum value in each case. The blue and green lines
) correspond to proportions of the remaining area covered by continents
and ocean . These results are based on 500 synthetic topography simulations
of an Earth-like planet ($H$ = 0.5, $\alpha$ = 1.9, $C_{1}$ = 0.1).
The red diamonds are for the Earth. \label{fig:Ocean/continent-relationship.-Si}}
\end{figure}

To estimate the properties of potential exoplanet surfaces, we conducted
a statistical analysis of oceans and continents obtained from 500
simulated multifractal topography fields at 1\textdegree{} spatial
resolution with the set of parameters obtained for the global estimates
on Earth ($H=0.5$, $\alpha=1.9$, $C_{1}=0.1$). In order to deal
with the notion of oceans and continents, one must first define the
sea/land cover. We define the sea level $s$, as a quantile of the
global topographic distribution. This definition simply means that
at quantile $s$, the sea level is such as $s$ is also the surface
proportion of the sea. For instance, (i) $s=0.5$ is the median altitude
and half of the planet is ocean covered , half by land; (ii) $s=0.9$
means that 90\% of the planet area is ocean covered and 10\% is land.
Oceans and continents are respectively defined as disconnected areas
located beneath or above the sea level $s$. We plotted on figure
\ref{fig:Example-of-180x360} a example of synthetic multifractal
topographies with varying ratio $s$.

On Figure \ref{fig:Ocean/continent-relationship.-Si} we plotted the
size of the largest continent and largest ocean as functions of $s$.
We summarized the 500 experiments by computing the average, standard
deviation and minimum/maximum. As one can see, the simulations produce
typically one large ocean or one large continent with a size close
to the maximum available area indicating that it is highly improbable
to obtain two disconnected large areas. However, at respectively very
small or very large values of $s$, the available area is split between
several small oceans (conversely, large $s$ and small continents).
Finally, we apply the same analysis on the particular case of Earth
based on ETOPO1 \citep{Amante2009} and use red diamonds to indicate
the size of the largest ocean and continents as functions of the terrestrial
value of s ($s\approx0.66$). The points are satisfyingly close to
those obtained by multifractal simulations supporting the accuracy
of the model. 

Following \citet{Dohm2015}, we investigate the interface between
ocean, atmosphere and land. From our results, on average the size
of the largest ocean or continent is always close to the maximum available
size (near the 90\% line). The congruent part of the surface covered
by ocean (or land) is split up into smaller but more numerous islands
(or lakes), as also observed on the Earth \citep{Downing_Theglobalabundance_LaO2006}.
There are some extreme cases, where the largest continent is very
small. Interestingly, this case happens more for small sea levels.
If $s=0.1$, the extreme case can even reach 25\%, meaning that the
largest ocean only covers 25\% of the ocean surface , 75\% are thus
covered by smaller lakes. The symmetric situation occurs for $s=0.9$
: the largest continent only covers 25\% of the land, 75\% are thus
covered by small islands. The Earth corresponds to the average situation
since all the major oceans are connected through the thermo-haline
circulation. From this study, we can exclude the situation of two
large unconnected oceans, representing a global sea surface > 50\%.
The same for two large unconnected continents, representing a global
sea surface > 50\%. As a summary, the interface between land and sea,
so important for habitability, can be statistically constrained by
this model.

\section{Conclusion}

Multifractal simulations on spheres are able to statistically reproduce
the morphology of planetary bodies, and even potentially small bodies
such asteroids and comets. In addition, it offers a wide field of
investigation for evaluating the role of the topography in exoplanet
signals , thanks to photometry and specular reflection, this is especially
true for transiting objects . The simulations will serve as a starting
point for future studies aimed at characterizing the overall photometric
response of unresolved rotating bodies. Our synthetic numerical topographies
can be integrated into the development of realistic exoplanet climate
simulations in different contexts by integrating the roles of clouds
and surface / atmosphere interactions. In particular, exoplanets in
gravitational lock are subjected to climatic instabilities \citep{Kite2011}.
In particular, our results suggest that it is statistically highly
unlikely to have two major united oceans on either side of the globe.
If the dark side is too cold and the sunny side too hot to allow the
presence of liquid water, the topography could contribute to creating
to a global glacier, continually moving the volatile elements from
the illuminated side to the dark side. This dynamic state should significantly
increase the presence of liquid water at the terminator with consequences
for habitability.

By construction the statistical properties of all our simulations
are isotropic. The procedure used can be modified to generate anisotropic
topographies but poses a number of technical problems that have not
yet been addressed. Anisotropy adds degrees of freedom that make the
problem more complex both in generation but also in determining parameters
on real data. To deal with this question, we should consider implementing
the formalism of generalized scale invariance (GSI, \citealp{Schertzer2011})
as a future work. 

We provide a 3D visualization of some examples with varying parameters
(https://data.ipsl.fr/exotopo/). In addition, a dataset of synthetic
spherical topographies can be downloaded by the reader (http://dx.doi.org/10.14768/20181024001.1) 

\section*{Acknowledgement}

We acknowledge support from the ``Institut National des Sciences
de l'Univers'' (INSU), the \textquotedbl{}Centre National de la Recherche
Scientifique\textquotedbl{} (CNRS) and \textquotedbl{}Centre National
d'Etudes Spatiales\textquotedbl{} (CNES) through the \textquotedbl{}Programme
National de Plan\'{e}tologie\textquotedbl{} and the \textquotedbl{}Programme
National de T\'{e}l\'{e}d\'{e}tection spatiale\textquotedbl{}, the MEX/OMEGA and
the MEX/PFS programs. We thank R. Orosei and the Assistant Editor
M. Hollis for their constructive reviews. We thank C. Marmo for the development of the 3D visualization tool and W. Pluriel for his contribution to the project. We also thank ESPRI-IPSL for hosting the data.

\bibliographystyle{agu04}

\end{document}